\documentclass[cameraready]{Interspeech}

\title{CoSTALA: Compositional Spatio-Temporal Audio-Language Alignment \\
via Multi-Grain Hierarchical Contrastive Learning}

\author[affiliation={1}]{Peiwei}{Ren}
\author[affiliation={2}]{Jinbo}{Hu}
\author[affiliation={3}]{Fang}{Kang}
\author[affiliation={1}]{Shan}{Liang}
\author[affiliation={4}]{Yin}{Cao}

\address{
    $^1$ Xi’an Jiaotong Liverpool University, China \\
    $^2$ MiLM Plus, Xiaomi Inc., China \\
    $^3$ Center for Machine Vision and Signal Analysis, University of Oulu, Finland \\
    $^4$ Institute of Acoustics, Chinese Academy of Sciences
}

\email{peiwei.ren25@student.xjtlu.edu.cn, hujinbo@xiaomi.com, \\
fang.s.kang@gmail.com, shan.liang@xjtlu.edu.cn, yin.k.cao@gmail.com}


\usepackage{comment}
\usepackage{amsmath}

\usepackage{microtype}
\usepackage{graphicx}
\usepackage{multirow}
\usepackage{booktabs}
\usepackage{amssymb}
\usepackage{tabularx}
\usepackage{cite}
\usepackage{hyperref}

\begin{document}

\maketitle

\begin{abstract}
Conventional audio language models (ALMs) have made significant progress in achieving alignment between auditory and textual representations, including recent explorations in spatial audio. However, in daily spatial scenarios, they still cannot effectively process multi-event audio sequences. Current approaches primarily rely on coarse-grained contrastive learning with global auditory and textual features, lacking the resolution to distinguish multiple sequential events. To overcome these limitations, we propose CoSTALA—a novel training paradigm that transitions from purely global alignment to fine-grained spatio-temporal reasoning. By constructing a multi-granularity hierarchical loss function system, we achieve explicit modeling of temporal dependencies, and successfully anchors individual acoustic events to preserve their semantic purity. Extensive experiments demonstrate that CoSTALA significantly establish a powerful new framework for spatio-temporal audio understanding.\footnote{Codes are available on: \url{https://github.com/Cell778/CoSTALA26.git}}

\end{abstract}

\begin{figure}[t]
    \centering
    \includegraphics[width=1.0\linewidth]{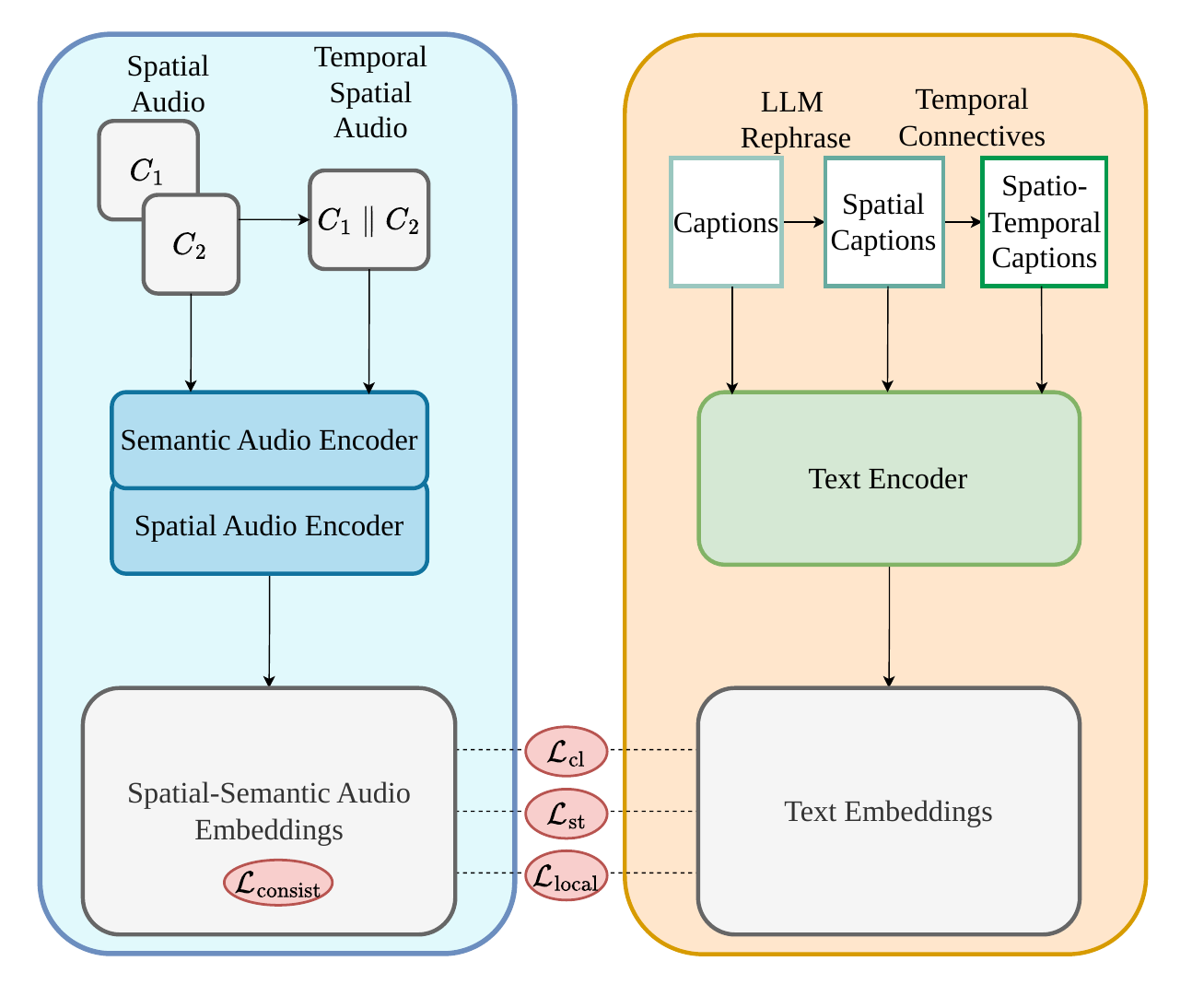}
    \caption{Overview of the CoSTALA framework for multi-grain spatio-temporal audio-language contrastive learning. The architecture is designed to process long-duration temporal spatial audio and corresponding spatio-temporal captions derived from isolated spatial events. A text encoder, a semantic audio encoder and a spatial audio encoder process these inputs to extract multi-level representations, which are jointly aligned through a suite of hierarchical loss functions. Specifically, a contrastive learning loss $\mathcal{L}_{\mathrm{cl}}$ and a local alignment loss $\mathcal{L}_{\mathrm{local}}$ establish multi-grain spatial audio-text mappings, while a spatio-temporal loss $\mathcal{L}_{\mathrm{st}}$ explicitly models chronological dependencies, and a feature consistency loss $\mathcal{L}_{\mathrm{consist}}$ preserves local semantic purity.}
    \vspace{-10pt}
    \label{fig:model}
\end{figure}

\section{Introduction}
Humans can naturally understand the appearance of objects through visual signals and describe them using natural language. Similarly, when a sound event reaches the human ears, the brain naturally interprets it and its direction of origin, prompting humans to describe it in language. For instance, a human perceives \textquotedblleft a dog is barking from the north, then a child is screaming from the south\textquotedblright\ through auditory signals, and then easily understands the sound event (\textquotedblleft dog barking\textquotedblright, \textquotedblleft child screaming\textquotedblright) and transfers them into semantic signals with spatial information (\textquotedblleft north\textquotedblright, \textquotedblleft south\textquotedblright). For machines to comprehend complex auditory environments, they must integrate auditory signals into linguistic signals. By applying linguistic understanding to complex sound signals, they can determine what sound events are occurring, where they are taking place, and their sequence of occurrence.

Contrastive learning \cite{Chen2020ASF} has emerged as a robust framework for learning aligned representations across distinct modalities. This efficacy was further validated in the vision-language domain by CLIP \cite{Radford2021LearningTV}. Building upon this foundational concept, Contrastive Language-Audio Pretraining (CLAP) \cite{10095889} has pioneered cross-modal alignment by mapping audio clips and text into a shared latent space. Subsequent studies have significantly broadened this paradigm. For instance, LAION-CLAP \cite{10095969} leverages massive datasets for auditory semantic enhancement, while T-CLAP \cite{10734763} refines the contrastive objective by introducing temporal negatives. To transcend the constraints of vanilla CLAP, recent innovations have targeted distinct dimensions: M2D-CLAP \cite{Niizumi2025M2DCLAPEG} enhances universal feature generalization via self-supervised learning; MGA-CLAP \cite{liAdvancingMultigrainedAlignment2024} shifts focus toward fine-grained alignment using discrete codebooks; and DRCap \cite{10890325} bridges the gap to generative tasks by integrating Large Language Models (LLMs) for zero-shot captioning.

Despite these advances, conventional CLAP frameworks remain predominantly anchored in static global alignment, primarily optimized for single-event or basic binaural audio-to-text mapping. Recent studies \cite{Devnani2024LearningSL,Zheng2024BATLT,Hu2025SALMSA, Yu2025TowardsMQ,Wilkinghoff2025DSpASTDR, Sakshi2025SPURAP,Sudarsanam2025TowardsSA,Tang2024CanLL} have extended this paradigm to spatial audio environments, showing promise in spatially-aware Question-Answering (QA) and text-queried tasks \cite{Zhao2024TextQueriedTS,Sudarsanam2025TowardsSA}. However, transitioning from static single-event representations to multi-event spatio-temporal sequences introduces significant theoretical challenges. As inputs expand into complex combinatorial spaces—where multiple acoustic events unfold across distinct spatial coordinates—relying solely on global representations becomes suboptimal. This coarse-grained approach imposes a severe information bottleneck by compressing dynamic, high-dimensional content into a single holistic vector. Such compression often induces \textquotedblleft context drift\textquotedblright, where the model averages out precise local semantics and consequently fails to resolve the distinct semantic identity of individual events.

To address this representation bottleneck, we argue that the structural correspondence between complex spatial audio streams and compositional natural language necessitates refined hierarchical representations rather than a single global projection. Therefore, we propose CoSTALA training paradigm, achieving a leap from purely global alignment to multi-granularity spatio-temporal reasoning. Beyond sequence-level contrastive learning, CoSTALA explicitly decouples and models the temporal dependencies of spatial events. By establishing local anchors via fine-grained alignment, we mitigate the risk of global feature collapse and preserve the semantic purity of individual acoustic events. Simultaneously, we introduce feature consistency and a 3-way spatio-temporal objective to strictly enforce combinatorial ordering in the latent space. By aligning features at both the localized event level and the global sequence level, our approach resolves temporal ambiguity and establishes a highly discriminative, granular foundation for complex spatial audio-language understanding.

To implement the proposed paradigm, the CoSTALA architecture explicitly maps these decoupled attributes to a shared latent space. We integrate a RoBERTa-based text encoder \cite{Liu2019RoBERTaAR} with a hierarchical audio backbone driven by HTSAT \cite{9746312}, as shown in Figure \ref{fig:model}. Inspired by the EINV2 dual-branch design \cite{9413473}, we strictly decouple pure acoustic semantic extraction from spatial localization. Crucially, to precisely capture the temporal sequence of acoustic events, we incorporate a dedicated Transformer-based \cite{Vaswani2017AttentionIA} temporal encoder equipped with Rotated Position Embedding (RoPE) \cite{Su2021RoFormerET}. By synergistically decoupling spatial-semantic extraction with multi-scale temporal modeling, CoSTALA effectively translates hierarchical training objectives into robust, unified spatio-temporal representations.

\section{Method}

\subsection{Spatial Audio-Text Dataset}
To obtain precise spatial audio-text pairs, we deploy a spatialized variant of the Clotho dataset \cite{9052990}. Specifically, we synthesize First-Order Ambisonics (FOA) audio samples by convolving original monophonic audio signals with simulated Spatial Room Impulse Responses (SRIRs) \cite{Hu2024PSELDNetsPN}.

We augment the original audio descriptions by leveraging a pre-trained Large Language Model (LLM) Qwen3-8B \cite{Yang2025Qwen3TR} to generate natural and coherent spatialized descriptions. Specifically, we discretize the continuous azimuth information of the audio into eight directions at 45-degree intervals (such as “southeast” and “northwest”). Using these spatial labels, we prompt the LLM to rewrite the original audio descriptions, seamlessly incorporating spatial context. 

To train and evaluate spatio-temporal reasoning, we construct a multi-event dataset. Positive anchors are synthesized by concatenating two distinct spatialized events $C_1^{\mathrm{loc1}}$ and $C_2^{\mathrm{loc2}}$ with zero overlap, while linking their captions via temporal connectives. To ensure fine-grained discrimination, we design two types of hard negatives per anchor \cite{10734763, Seki2025SpatialCLAPLS, Xie2023EnhanceTR}: (1) \textbf{Temporal Negatives}, created by reversing the chronological order ($C_2^{\mathrm{loc2}}\parallel C_1^{\mathrm{loc1}}$); and (2) \textbf{Spatial Negatives}, formed by swapping spatial rendering locations ($C_1^{\mathrm{loc2}}\parallel C_2^{\mathrm{loc1}}$). Matching negative captions are procedurally generated. The resulting dataset comprises 375 hours of audio distributed across 30,000 training samples (equally split among positives and both types of negatives) and 9,000 evaluation samples.

\subsection{Embedding Structure}
\vspace{-10pt}
\begin{table}[htbp]
\centering
\caption{Summary of key embeddings used in CoSTALA.}
\label{tab:embeddings}
\renewcommand{\arraystretch}{1.2}
\begin{tabular}{clp{4.5cm}}
\toprule
\textbf{Modality} & \textbf{Notation} & \textbf{Description} \\
\midrule
\multirow{3}{*}{\textbf{Text}} 
& $T_{\mathrm{sem}}$ & Semantic embeddings derived from concatenated captions. \\
& $T_{\mathrm{spa}}^{\mathrm{c}i}$ & Spatial embeddings of directional captions for chunk $i$. \\
& $T_{\mathrm{st}}$ & Unified spatio-temporal embeddings of spatial multi-event captions. \\
\midrule
\multirow{5}{*}{\textbf{Audio}} 
& $E_{\mathrm{hard}}^{\mathrm{c}i}$ & Joint spatio-semantic representations from isolated event $C_i$. \\
& $E_{\mathrm{soft}}^{\mathrm{c}i}$ & Joint spatio-semantic representations from temporally sliced global maps. \\
& $E_{\mathrm{temp}}$ & Temporal representations capturing sequential dependencies. \\
& $E_{\mathrm{global}}$ & Macroscopic joint representations of concatenated audio $C_1\parallel C_2$. \\
& $E_{\mathrm{st}}$ &  Spatio-temporal representations. \\
\bottomrule
\end{tabular}
\end{table}

\begin{figure*}
    \centering
    \vspace{-20pt}
    \includegraphics[width=0.95\textwidth]{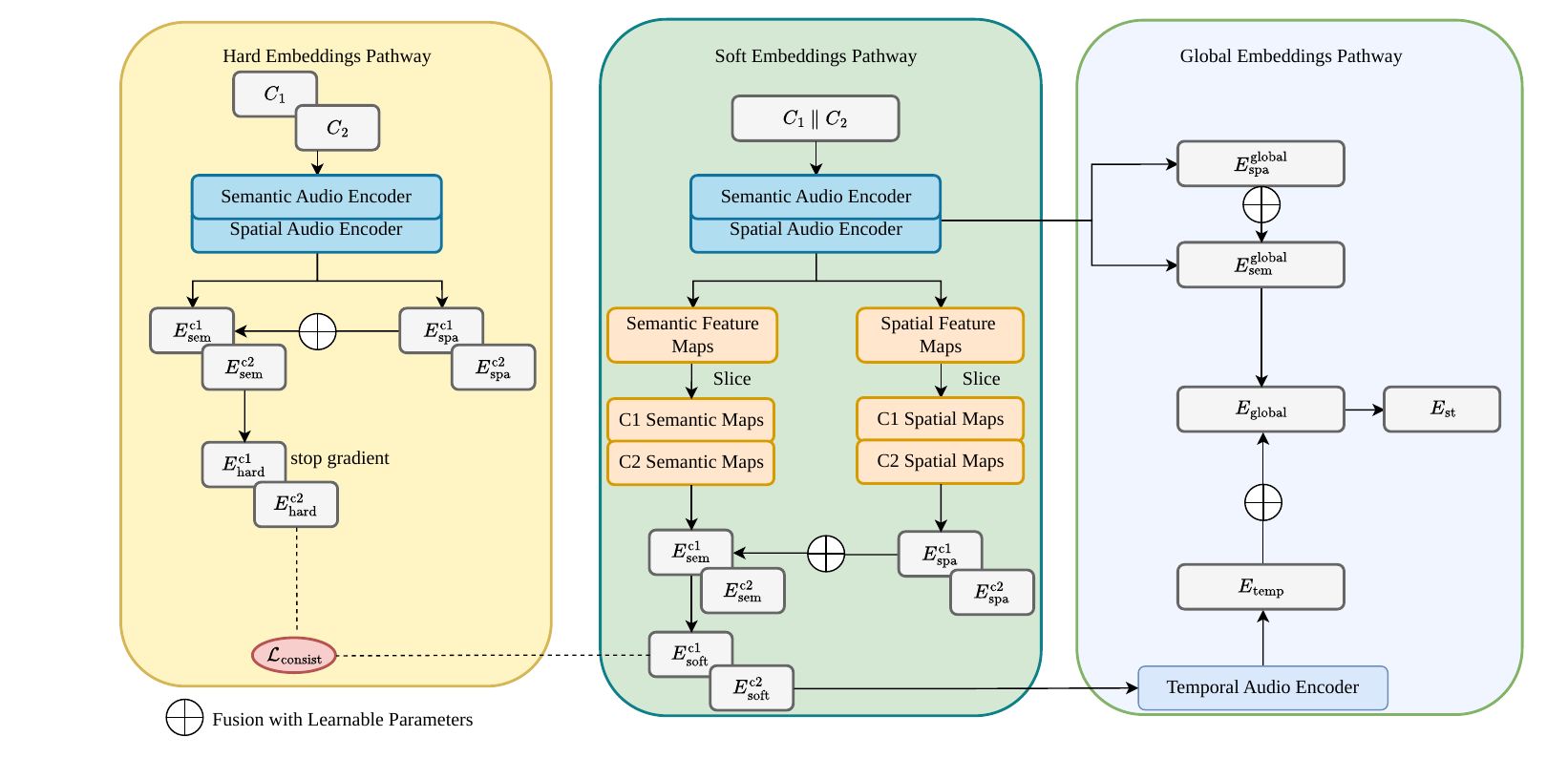}
    \caption{Detailed audio pipeline. A shared encoder processes isolated and composed events into hard ($E_{\mathrm{hard}}$) and soft ($E_{\mathrm{soft}}$) representations. $E_{\mathrm{hard}}$ anchors consistency learning ($\mathcal{L}_{\mathrm{consist}}$). A temporal encoder extracts sequential features ($E_{\mathrm{temp}}$) from $E_{\mathrm{soft}}$, which are fused with $E_{\mathrm{global}}$ to yield $E_{\mathrm{st}}$.}
    \label{fig:figure2}
\end{figure*}

The text encoder generates three distinct levels of textual representations: semantic embeddings $T_{\mathrm{sem}}$ derived from the concatenation of original captions; spatial embeddings $T_{\mathrm{spa}}^{\mathrm{ci}}$ extracted from LLM-augmented directional captions; and unified spatio-temporal embeddings $T_{\mathrm{st}}$ derived from multi-event descriptions joined by temporal connectives.

As illustrated in Figure \ref{fig:figure2}, our audio processing pipeline comprises three distinct pathways, leveraging both a semantic audio encoder and a spatial audio encoder.

\smallskip
\noindent\textbf{Hard Embeddings Pathway}. We process isolated spatial events $[C_1,C_2]$ to derive their semantic and spatial features, denoted as $[E_{\mathrm{sem}}^{\mathrm{c1}}, E_{\mathrm{sem}}^{\mathrm{c2}}]$, and $[E_{\mathrm{spa}}^{\mathrm{c1}}, E_{\mathrm{spa}}^{\mathrm{c2}}]$. These modality-specific representations are then fused via a weighted addition mechanism parameterized by a learnable weight vector $\mathbf{w} \in \mathbb{R}^{512}$ to yield joint spatio-semantic embeddings $[E_{\mathrm{hard}}^{\mathrm{c1}}, E_{\mathrm{hard}}^{\mathrm{c2}}]$.

\smallskip
\noindent\textbf{Soft Embeddings Pathway}. To capture fine-grained temporal dynamics, we input the concatenated audio $C_1\parallel C_2$ to explicitly extract global semantic and spatial feature maps. These maps are subsequently partitioned into chunk-level feature maps based on original segment durations. By employing the same fusion mechanism, we obtain the soft-sliced representations $[E_{\mathrm{soft}}^{\mathrm{c1}}, E_{\mathrm{soft}}^{\mathrm{c2}}]$. Finally, this sequence is fed into the temporal audio encoder to capture chronological dependencies, yielding the temporal embedding, $E_{\mathrm{temp}}$.

\smallskip
\noindent \textbf{Global Embeddings Pathway}. We process the concatenated audio sequence $C_1 \parallel C_2$, to extract global-level semantic representations, $E_{\mathrm{sem}}^{\mathrm{global}}$, and spatial representations, $E_{\mathrm{spa}}^{\mathrm{global}}$. These features are subsequently integrated into a macroscopic joint embedding, $E_{\mathrm{global}}$. We then employ a learnable scalar $\alpha$ to fuse $E_{\mathrm{temp}}$ with this macroscopic joint embedding, $E_{\mathrm{global}}$, via a residual connection, yielding the spatio-temporal audio representation, $E_{\mathrm{st}}$.

A comprehensive summary of these embeddings is provided in Table \ref{tab:embeddings}.

\subsection{Multi-Grain hierarchical Loss Functions}
To facilitate stable optimization during long-sequence learning, we propose a comprehensive suite of hierarchical loss functions. 

\smallskip
\noindent \textbf{Contrastive learning loss.} First, to establish a foundational audio-language alignment, we employ a contrastive learning objective \cite{10095889} across all samples, where each positive anchor is grouped with its corresponding temporal and spatial hard negatives within the same batch. We define two distinct alignment pairs: a purely semantic pair $[E_{\mathrm{sem}}, T_{\mathrm{sem}}]$ and a joint macroscopic pair $[E_{\mathrm{global}}, T_{\mathrm{st}}]$. Denoting the contrastive losses for these two pairs as $\mathcal{L}_{\mathrm{cl}}^{\mathrm{sem}}$ and $\mathcal{L}_{\mathrm{cl}}^{\mathrm{st}}$ respectively, the total contrastive loss $L_{\mathrm{cl}}$ is formulated as the sum of these two components.

\smallskip
\noindent \textbf{3-way spatio-temporal loss.} To discriminate combinatorial nuances beyond macroscopic patterns, we formulate a 3-way contrastive loss \cite{10734763, Seki2025SpatialCLAPLS} leveraging the explicitly modeled spatio-temporal embeddings, $E_{\mathrm{st}}$ and $T_{\mathrm{st}}$. For a batch of $N$ samples, the text-to-audio objective compels the model to differentiate between chronological and spatial errors:

\begin{equation}
\mathcal{L}_{\mathrm{st}}^{\mathrm{text} \to \mathrm{audio}} = -\frac{1}{N}\sum_{i=1}^{N} \log \left( \frac{\exp( (T_{\mathrm{st}}^{(i)} \cdot E_{\mathrm{st}}^{(i)})/\tau)}{\mathcal{Z}_{i}} \right)
\end{equation}
where $\tau$ is a learnable temperature parameter, and the denominator $\mathcal{Z}_{i}$ encapsulates the positive anchor pair alongside the two types of audio hard negatives:

\begin{equation}
\begin{split}
\mathcal{Z}_{i} &= \exp( (T_{\mathrm{st}}^{(i)} \cdot E_{\mathrm{st}}^{(i)})/\tau) + \exp( (T_{\mathrm{st}}^{(i)} \cdot \tilde{E}_{\mathrm{st}}^{(i)})/\tau) \\
&\quad + \exp((T_{\mathrm{st}}^{(i)} \cdot \dot{E}_{\mathrm{st}}^{(i)})/\tau)
\end{split}
\end{equation}
Here, $\tilde{E}_{\mathrm{st}}^{(i)}$ and $\dot{E}_{\mathrm{st}}^{(i)}$ represent the temporal and spatial negatives. The objective of $\mathcal{L}_{\mathrm{st}}^{\mathrm{text} \to \mathrm{audio}}$ is to force the model to explicitly discriminate between chronological errors and spatial localization errors, using the positive text embedding, $T_{\mathrm{st}}^{(i)}$, as the anchor. 

Analogously, we formulate the symmetric audio-to-text loss $\mathcal{L}_{\mathrm{st}}^{\mathrm{audio} \to \mathrm{text}}$ by using the anchor audio embedding, $E_{\mathrm{st}}^{(i)}$, with negative text embeddings, $\tilde{T}_{\mathrm{st}}^{(i)}$ and $\dot{T}_{\mathrm{st}}^{(i)}$. Finally, the total 3-way spatio-temporal loss $L_{\mathrm{st}}$ is formulated as the sum of these two symmetric components.

\begin{table*}[t]
\centering
\vspace{-10pt}
\caption{Ablation study of the proposed hierarchical loss functions on spatial retrieval performance. We report Recall@$K$ ($K \in \{1, 5, 10\}$) in percentage (\%) for both Text-to-Audio and Audio-to-Text directions. The left section evaluates the global spatio-temporal features, while the right section assesses the pure auditory semantic representations.}
\label{tab:main_ablation}
\renewcommand{\arraystretch}{1.15}
\resizebox{\textwidth}{!}{
\begin{tabular}{l ccc ccc ccc ccc}
\toprule
\multirow{3}{*}{\textbf{Model / Loss Config}} & \multicolumn{6}{c}{\textbf{Global Spatio-Temporal Retrieval}} & \multicolumn{6}{c}{\textbf{Semantic-Only Retrieval}} \\
\cmidrule(lr){2-7} \cmidrule(lr){8-13}
& \multicolumn{3}{c}{Text-to-Audio} & \multicolumn{3}{c}{Audio-to-Text} & \multicolumn{3}{c}{Text-to-Audio} & \multicolumn{3}{c}{Audio-to-Text} \\
\cmidrule(lr){2-4} \cmidrule(lr){5-7} \cmidrule(lr){8-10} \cmidrule(lr){11-13}
& R@1 & R@5 & R@10 & R@1 & R@5 & R@10 & R@1 & R@5 & R@10 & R@1 & R@5 & R@10 \\
\midrule
SALM \cite{Hu2025SALMSA} & 4.77 & 14.37 & 21.27 & 4.57 & 14.28 & 21.23 & - & - & - & - & - & - \\
T-CLAP \cite{10734763} & 5.66 & 16.84 & 24.88 & 5.92 & 16.71 & 24.71 & - & - & - & - & - & - \\
\midrule
CoSTALA ($\mathcal{L}_{\mathrm{cl}}$) 
& 5.84 & 17.68 & 26.07 & 6.58 & 18.49 & 26.01 & 1.87 & 7.29 & 12.44 & 1.61 & 5.23 & 8.30 \\
CoSTALA ($\mathcal{L}_{\mathrm{cl}} + \mathcal{L}_{\mathrm{st}}$) 
& 7.16 & 17.45 & 24.52 & 6.96 & 17.78 & 23.84 & 1.70 & 6.95 & 11.57 & 0.91 & 3.96 & 6.81 \\
CoSTALA ($\mathcal{L}_{\mathrm{cl}} + \mathcal{L}_{\mathrm{st}} + \mathcal{L}_{\mathrm{local}}$) 
& 6.98 & 17.99 & 25.77 & 7.27 & 18.58 & 25.93 & 2.10 & 8.01 & 13.02 & 1.21 & 2.99 & 4.60 \\
CoSTALA ($\mathcal{L}_{\mathrm{cl}} + \mathcal{L}_{\mathrm{st}} + \mathcal{L}_{\mathrm{consist}}$) 
& 6.82 & 17.39 & 24.80 & 6.21 & 16.29 & 23.22 & 1.65 & 6.54 & 11.01 & 1.13 & 4.08 & 6.82 \\
CoSTALA ($\mathcal{L}_{\mathrm{cl}} + \mathcal{L}_{\mathrm{st}} + \mathcal{L}_{\mathrm{local}} + \mathcal{L}_{\mathrm{consist}}$)
& \textbf{8.10} & \textbf{19.86} & \textbf{27.68} & \textbf{8.16} & \textbf{20.49} & \textbf{27.08} & 2.11 & 8.25 & 13.60 & 1.07 & 3.74 & 7.26 \\
\bottomrule
\end{tabular}
}
\end{table*}

\smallskip
\noindent \textbf{Local alignment loss.} To mitigate the erosion of fine-grained details inherent in global alignment, we incorporate a local alignment objective. By explicitly aligning the independent hard-encoded audio chunks, $E_{\mathrm{hard}}^{\mathrm{c}i}$, with their corresponding spatial captions, $T_{\mathrm{spa}}^{\mathrm{c}i}$, we establish robust semantic anchors for individual acoustic events. This dense supervision prevents local features from being submerged by the global sequence context, thereby maintaining high semantic purity. For $i \in \{1, 2\}$, the local loss is formulated via the standard InfoNCE objective:

\begin{equation}
\mathcal{L}_{\mathrm{local}} = \sum_{i=1}^{2} \mathcal{L}_{\mathrm{InfoNCE}}(E_{\mathrm{hard}}^{\mathrm{c}i}, T_{\mathrm{spa}}^{\mathrm{c}i})
\end{equation}

\smallskip
\noindent \textbf{Feature consistency loss.} When modeling continuous audio streams, localized representations derived from a global sequence are inherently susceptible to temporal entanglement and semantic drift. To mitigate these effects, we introduce a feature consistency loss $\mathcal{L}_{\mathrm{consist}}$. By leveraging the independently encoded hard chunks, $E_{\mathrm{hard}}^{\mathrm{c}i}$, as anchors, we enforce a Mean Squared Error (MSE) constraint on the corresponding soft embeddings $E_{\mathrm{soft}}^{\mathrm{c}i}$. Crucially, a stop-gradient operation is applied to the hard anchors to prevent representational collapse and stabilize training. This explicit regularization ensures that soft representations preserve high semantic fidelity within complex global contexts. For $i \in \{1, 2\}$, the objective is formulated as:

\begin{equation}
\mathcal{L}_{\mathrm{consist}} = \sum_{i=1}^{2} \lVert E_{\mathrm{soft}}^{\mathrm{c}i} - \mathrm{sg}(E_{\mathrm{hard}}^{\mathrm{c}i}) \rVert_{2}^{2}
\end{equation}

\smallskip
\noindent \textbf{Overall training objective.} To achieve comprehensive alignment across all granularities, we jointly optimize the multi-grain spatio-temporal representations and temporal reasoning capabilities. The overall training objective is formulated as a weighted linear combination of the aforementioned constraints:

\begin{equation}
\mathcal{L}_{\mathrm{total}} = \sum_{k \in \mathcal{K}} \lambda_{k} \mathcal{L}_{k}
\end{equation}
where the task set $\mathcal{K} = \{\mathrm{cl}, \mathrm{st}, \mathrm{local}, \mathrm{consist}\}$, and $\lambda_{k}$ are empirical hyperparameters scaling the relative contribution of each specific objective.

\section{Experiments}
\subsection{Experimental Setup}
Synthetic spatial audio signals are sampled at 24 kHz. We extract 64-dimensional log-mel spectrograms and intensity vectors from the four-channel First-Order Ambisonics (FOA) signals using a 1,024-point Hanning window with a 240-point hop size. The framework is trained for 15 epochs using the AdamW optimizer with a peak learning rate of $10^{-4}$, incorporating a 3-epoch linear warm-up phase followed by a cosine annealing schedule. 

Following SALM \cite{Hu2025SALMSA}, we evaluate bi-directional spatial retrieval (i.e., audio-to-text and text-to-audio) using the Recall@$K$ metric ($K \in \{1, 5, 10\}$). The retrieval performance is computed based on the cosine similarity between the macroscopic joint audio embeddings, $E_{\mathrm{global}}$, and the corresponding text embeddings, $T_{\mathrm{st}}$.

\subsection{Spatial Retrieval Performance}
Table \ref{tab:main_ablation} presents a comprehensive comparative analysis and ablation study. We benchmark our framework against two representative ALMs, SALM and T-CLAP, which exemplify distinct paradigms in current audio-language modeling. Specifically, SALM emphasizes the global alignment of spatial audio understanding, but is fundamentally constrained by its coarse-grained global alignment. Conversely, while T-CLAP specializes in temporal multi-event reasoning, it is restricted to a two-channel auditory environment. Our proposed CoSTALA bridges these gaps by transitioning from global alignment to hierarchical spatio-temporal reasoning. 

CoSTALA (with only $\mathcal{L}_{\mathrm{cl}}$) already outperforms established ALMs such as SALM and T-CLAP by leveraging global in-batch spatio-temporal negative samples. Unlike SALM, which lacks explicit hard negative contrasting, and T-CLAP, whose 2-way loss solely targets chronological errors while ignoring spatial dynamics, $\mathcal{L}_{\mathrm{cl}}$ performs global alignment within a unified negative pool encompassing both spatially and temporally inconsistent samples. This large-scale contrastive setting enables the model to learn more discriminative global representation. 

Building upon this global pool discrimination, the introduction of the 3-way spatio-temporal loss, $\mathcal{L}_{\mathrm{st}}$, yields a significant increase in R@1. This suggests that its rigorous decision boundary successfully suppresses partially similar candidates (e.g., incorrect event ordering), and isolates the precise spatio-temporal match. Crucially, our ablation study demonstrates a potent synergy between local alignment loss, $\mathcal{L}_{\mathrm{local}}$, and feature consistency loss, $\mathcal{L}_{\mathrm{consist}}$. Notably, applying $\mathcal{L}_{\mathrm{consist}}$ without $\mathcal{L}_{\mathrm{local}}$ leads to performance degradation. However, when integrated into the full framework, the model achieves the peak performance. This vindicates our core hypothesis regarding feature collapse: only by simultaneously enforcing local grounding, $\mathcal{L}_{\mathrm{local}}$, and feature consistency, $\mathcal{L}_{\mathrm{consist}}$, can the model effectively counteract the overwhelming global context, and effectively mitigate the representational collapse inherent in long-sequence audio learning.

To further validate the indispensability of the spatial pathway, we conduct a zero-shot retrieval test pairing the purely semantic audio representation, $E_{\mathrm{sem}}$, with the complex spatio-temporal text representation, $T_{\mathrm{st}}$. As shown in the right section of Table \ref{tab:main_ablation}, relying solely on semantic audio representation leads to a catastrophic performance decline compared to the joint representation. This empirical finding rigorously demonstrates that while semantic modeling can identify acoustic entities, it inherently lacks the spatial granularity required to disentangle complex multi-event scenarios.

\section{Conclusion}
In this paper, we introduced a novel training paradigm Compositional Spatio-Temporal Audio-Language Alignment (CoSTALA) for the comprehensive understanding of long-duration, multi-event spatial audio sequences. By integrating explicit temporal modeling with a suite of multi-grain hierarchical losses, our approach effectively overcomes the fundamental limitations of existing Audio-Language Models in processing extended spatial contexts. Ultimately, this work establishes a robust and fine-grained foundational framework, paving the way for future spatio-temporal audio-language research.

\section{Generative AI Use Disclosure}
The generative AI is used for reviewing grammar, checking format, and polishing manuscript. 

\section{Acknowledgements}
This work was supported by the Research Development Fund (RDF) of Xi'an Jiaotong-Liverpool University under Grant No. RDF-22-01-084.

\bibliographystyle{IEEEtran}
\bibliography{mybib}

\end{document}